# Defect migration and phase transformations in 2D iron chloride inside bilayer graphene

Qiunan Liu[1], Haiming Sun[1], Yung-Chang Lin[1,2]*, Mahdi Ghorbani-Asl[3], Silvan Kretschmer[3], Chi-Chun Cheng[4], Po-Wen Chiu[4,5], Hiroki Ago[6,7], Arkady V. Krasheninnikov[3,1]*, Kazu Suenaga[1,2]*

[1]The Institute of Scientific and Industrial Research (ISIR-SANKEN), Osaka University, Osaka 567-0047, Japan

[2]Nanomaterials Research Institute, National Institute of Advanced Industrial Science and Technology (AIST), Tsukuba 305-8565, Japan

[3]Institute of Ion Beam Physics and Materials Research, Helmholtz-Zentrum Dresden-Rossendorf, 01328 Dresden, Germany

[4]Department of Electrical Engineering, National Tsing Hua University, Hsinchu 30013, Taiwan

[5]Institute of Atomic and Molecular Sciences, Academia Sinica, Taipei, 10617, Taiwan

[6]Global Innovation Center (GIC), Kyushu University, Fukuoka 816-8580, Japan

[7]Interdisciplinary Graduate School of Engineering Sciences, Kyushu University, Fukuoka 816-8580, Japan

## Abstract

The intercalation of metal chlorides, and particularly iron chlorides, into graphitic carbon structures has recently received lots of attention, as it can not only protect this two-dimensional (2D) magnetic system from the effects of the environment, but also substantially alter the magnetic, electronic, and optical properties of both intercalant and host material. At the same time, the intercalation can result in the formation of structural defects, or defects can appear under external stimuli, which can affect materials performance. These aspects have received so far little attention in the dedicated experiments. In this study, we investigate the behavior of atomic-scale defects in iron chlorides intercalated into bilayer graphene (BLG) by using scanning transmission electron microscopy (STEM) and first-principles calculations. We observe transformations between the FeCl$_2$ and FeCl$_3$ phases and elucidate the role of defects in the transformations. Specifically, three types of defects are identified: Fe vacancies in FeCl$_2$ domains, Fe adatoms and interstitials in FeCl$_3$ domains, each exhibiting distinct dynamic behaviors. We also observed a crystalline phase with an unusual stoichiometry of Fe$_5$Cl$_{18}$ which has not been reported before. Our findings not only advance the understanding of intercalation mechanism of 2D materials but also highlight the profound impact of atomic-scale defects on their properties and potential technological applications.

**Keywords:** Intercalation, STEM, iron chloride, phase transformation, bilayer graphene

**Corresponding authors:** yc-lin@aist.go.jp, a.krasheninnikov@hzdr.de, suenaga-kazu@sanken.osaka-u.ac.jp

The intercalation of metal chlorides into sp$^2$-hybridized carbon systems has been the subject of scientific investigations since 1932.[1] These materials are among the most commonly used intercalants into graphite as well as few-layer and bilayer graphene (BLG).[2] Recent experiments unveiled complexities in the intercalated structures that often diverge from initial theoretical expectations, see Ref.[3] for an overview. For instance, experiments on intercalation of iron (III) chloride (FeCl$_3$) in BLG revealed its coexistence with iron (II) chlorides (FeCl$_2$),[4] aluminum chloride (AlCl$_3$) was found to form unexpected polymorphic phases upon intercalation,[5] along with the reduction of molybdenum pentachloride (MoCl$_5$) to molybdenum trichloride (MoCl$_3$).[6] Moreover, intercalated alkali metals form bilayers and multilayers within BLG,[7, 8] contrary to expected monolayer structures. These findings highlight the intricate nature of intercalation mechanism and underscore the importance of direct visualization of the intercalated structures using atomic resolution techniques to deepen the understanding of the intercalation process and characteristics of the intercalated systems.

The intercalated materials frequently have atomic-scale defects which can strongly influence their properties, often leading to the degradation of its performance, but defects can also be beneficial. They generally govern the mechanical strength,[9] electrical conductivity,[10] and catalytic activity,[11,12] and can also facilitate phase transformations.[13-17] In that context, the migration of point defects, that is the motion of atomic vacancies and/or interstitial atoms within the crystal lattice has garnered significant attention.[18-20] The reduced dimensionality of 2D materials enhances the impact of defects as well as their visibility, allowing for detailed investigation using advanced microscopy techniques.

In this study, we investigate the intercalation of $FeCl_3$ into BLG by using low-voltage aberration-corrected scanning transmission electron microscopy (STEM). Our results confirm the coexistence of multiple $FeCl_x$ phases of iron chloride within BLG, and the phase transformations between them were observed *in situ* at the atomic scale. By tracking the behaviors of individual atomic defects, we have successfully elucidated the atomic mechanisms of defect-related phase transformations in this system.

## Results and Discussion

### Nanostructures and interfaces of intercalated iron chlorides in BLG.

The samples intercalated with $FeCl_3$ in BLG were prepared using the chemical vapor transport (CVT) method.[5,6] BLG was initially transferred onto a TEM grid and placed with $FeCl_3$ powders in a vacuum-sealed glass tube ($P < 10^{-3}$ Pa). The intercalation process involved heating the sealed glass tube in a box furnace at 250 °C for 12 hours. After the process, the tubes were carefully opened in an argon-filled glove box to prevent exposure of the samples to the atmosphere and potential oxidation. They were then transferred to an ultra-high vacuum (UHV) TEM using a JEOL vacuum transfer holder to ensure the integrity of the sample during the TEM investigations. To reduce beam damage during the observation, the STEM experiment was performed using a low-voltage (60kV) electron source. Also the vacuum level in the specimen chamber reached approximately $5\times10^{-7}$ Pa. At this level of vacuum, potential chemical etching of the graphene by the 60kV electron beam can be considered negligible.[21] Prior to the intercalation experiment, the BLG surface was thoroughly cleaned through thermal annealing at 250°C in air for 30 minutes, followed by annealing in an Ar/$H_2$ environment for 1 hour.[22] This process ensured that the BLG served as a clean and highly effective protective layer for observing the intercalants by electron beam. Most importantly, the probe current was reduced to 10 pA, which minimized beam-induced damage and allowed for stable structural observations.

The intercalated materials were found to not only contain $FeCl_3$, but also $FeCl_2$, with the two phases often separated by sharp boundaries, as depicted in the annular dark field (ADF) image, **Figure 1a**. In STEM ADF imaging, contrast follows Z-contrast, where heavier atoms appear brighter and lighter atoms appear darker.[23] The atomic numbers of Fe, Cl, and C are 26, 17, and 6, respectively. In the ADF image shown in Figure 1a, the BLG structure is barely visible; however, its presence is clearly confirmed by the FFT spots marked by red and yellow circles in Figure S1b. The twist angle between the two graphene layers is approximately 21°. A corresponding atomic model and STEM image simulation are presented in Figure S1c,d. The constructed theoretical model of $FeCl_3$ sandwiched between graphene bilayer, after fully relaxation, exhibits a thickness of approximately 9.8 Å (Figure S2a). Experimentally, the thickness derived by EELS data

(zero-loss spectrum) is estimated to be ~1 nm, which aligns closely with the theoretical value (Figure S2b-d). Although the atomic model indicates that the BLG and $FeCl_3$ form a superlattice structure, the BLG is difficult to discern in the simulated STEM image due to the weak contrast of carbon compared to Fe and Cl. Interestingly, one can see that the lattice imperfections in the $FeCl_2$ and $FeCl_3$ structures are distinctly different. In the $FeCl_3$ lattice, extra Fe atoms are the predominating defects, manifesting as interstitials in the lattice and adatoms on the $FeCl_3$ surface appearing as brighter contrast atoms on top of Fe sites, indicated by the red arrow in Figure 1a. To investigate the structure of adatoms on $FeCl_3$, we conducted STEM simulations based on theoretical models of Fe, Cl, and C adatoms on $FeCl_3$, as shown in Figure S3(a-h). By carefully comparing the intensity profiles of the three simulated ADF images with the experimental data, we determined that the simulated profile of Fe adatoms aligns well with the experimental results (Figure S3i). Conversely, in the $FeCl_2$ lattice, Fe vacancies can be identified as the most prolific defects by the darker contrast at the Fe sites, indicated by the white arrow. We did not observe any adsorbed atoms on the surface of $FeCl_2$. A sharp and seamless interface forms between $FeCl_2$ and $FeCl_3$ when they are aligned with the same orientation. However, if the two intercalated iron chloride domains are joined at an angle, a domain boundary forms, which consists of heptagonal structures, as illustrated in Figure 1b-1d. The domain boundary can be deformed under electron beam irradiation, where the displacement of heptagons perpendicular to the grain boundary occurs due to the local migration of Fe atoms, as can be observed in Figure 1e-1g. Figure 1h and 1i present the atomic models of a Fe vacancy in $FeCl_2$ and Fe adatom in $FeCl_3$, respectively. Our DFT calculations indicate that the adsorption of Fe adatoms on Fe sites is energetically more favorable than on hollow sites, as observed in the $FeCl_3$ configuration (Figure 1j and S4). This occurs because Fe adatoms become incorporated into the lattice, forming a relatively stable dumbbell-like structure. The dumbbell-like Fe adatoms demonstrate substantial stability under 60 kV electron beam irradiation and preferentially migrate to another Fe site (Figure 1k and Movie S1), forming a new dumbbell, rather than jumping to the hollow site or creating Fe interstitials in $FeCl_3$. In contrast, the migration behavior of Fe interstitials in $FeCl_3$ domains and Fe vacancies in $FeCl_2$ domains differ significantly. These differences in defect dynamics and their influence on the material structural evolution will be discussed in detail in the following sections.

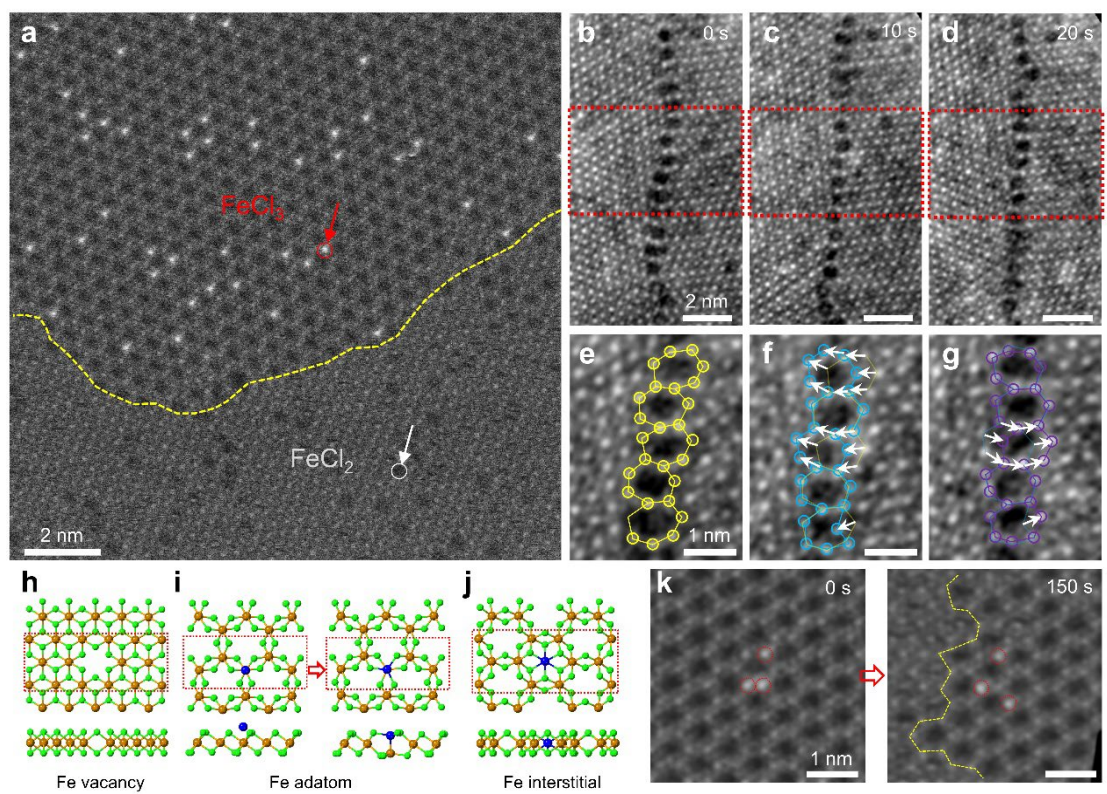

**Figure 1. Structural defects and interfaces of iron chlorides in BLG.** (a) ADF image showing the sharp interface between $FeCl_3$ and $FeCl_2$ domains. The interface is highlighted by the yellow dashed line. A Fe adatom is indicated by a red arrow within $FeCl_3$ domain, while a Fe vacancy is indicated by the white arrow within the $FeCl_2$ domain. (b-d) Consecutive ADF images depicting the grain boundary between two $FeCl_2$ domains that are joined at an angle of 30°. (e-g) Magnified ADF images from the red dotted boxes in (b-d). Colored heptagons and circles highlight the formation of the grain

boundary. The migration of Fe atoms at the boundary is indicated by white arrows. (h) Atomic model of a Fe vacancy in FeCl$_2$, top and side views. The red dotted boxes indicate the regions selected to present the cross-sectional structure. (i) Atomic model of a Fe adatom in FeCl$_3$ transforming into a dumbbell-like structure. (j) Atomic model of a Fe interstitial in FeCl$_3$. (k) Sequential ADF images of the migration of dumbbell-like Fe adatoms in FeCl$_3$.

## Migration of Fe vacancy in FeCl$_2$ and Fe interstitial in FeCl$_3$

Monolayer FeCl$_2$ crystallizes in the trigonal P-3m1 space group, with a structure similar to the 1T phase of transition metal dichalcogenides (TMDCs) such as MoS$_2$ and WS$_2$. In TMDCs, the mobility of chalcogen vacancies is much higher as compared to that of metal vacancies or substitutional dopants,[24, 25] leading to such phenomena as formation of three-fold rotational defects[26] or the migration of 5-7 defects.[27] In contrast, FeCl$_2$ often exhibits lattice imperfections such as single or multiple Fe vacancies. Unlike the relatively immobile metal vacancies in TMDC, Fe defects in FeCl$_2$ demonstrate considerable mobility under electron beam irradiation.

**Figure 2a** presents a sequence of ADF images that illustrate the migration process of a single Fe vacancy in FeCl$_2$. In Figure 2a(i), the center of the observed area contains a single Fe vacancy, marked by a red dashed circle. Under electron beam irradiation, this Fe vacancy gradually migrates towards the bottom right position after 16 scanning images (approximately 170 sec; see Movie S2). This observation indicates that Fe vacancies in FeCl$_2$ can exchange positions with the nearest neighboring Fe atoms. Figure 2b provides a model of the Fe vacancy migration path, showing that the minimum migration distance of the Fe vacancy corresponds to the lattice constant $a=b\approx3.5$ Å. The calculated migration barrier is 1.4 eV, as depicted in Figure 2e.

In the case of FeCl$_3$, it crystallizes in the trigonal P312 space group and has a structure similar to aluminum trichloride ($AlCl_3$). The defect structure in $FeCl_3$ contrasts with that of $FeCl_2$, involving Fe interstitials, where extra Fe atoms occupy vacant sites within the $FeCl_3$ framework. Figure 2c shows a sequence of ADF images depicting the migration of a Fe interstitial in $FeCl_3$ (see Movie S3). The Fe interstitial, marked by a cyan dashed circle, moves within the Fe atomic plane to fill in the nearest vacant Fe sites. The minimum migration path for the isolated Fe interstitial follows the lattice constant of $FeCl_3$, with a migration distance of *a+b* = 6.2 Å, as indicated by the blue arrows in the corresponding migration path model in Figure 2d. The migration barrier of the Fe interstitial in $FeCl_3$ is 1.8 eV (Figure 2f). Although the calculated migration barrier of interstitials in $FeCl_3$ is slightly higher than that of vacancies in $FeCl_2$, our STEM experiments reveal that Fe interstitials in $FeCl_3$ exhibit higher mobility than Fe vacancies. However, the migration of the defects during e-beam scanning can be different from thermally activated diffusion as it is accompanied by chlorine atom displacement, which directly influences structural transformations. In principle, sufficient momentum transfer from the electron beam to the target material can induce lateral atomic migration, as demonstrated in graphene doped with impurities using low-voltage STEM.[28] This technique represents a critical advancement for nanotechnology, particularly if the electron probe can be precisely positioned at specific sites based on guidance from DFT calculations to transfer kinetic energies for Fe interstitials and vacancies migration.

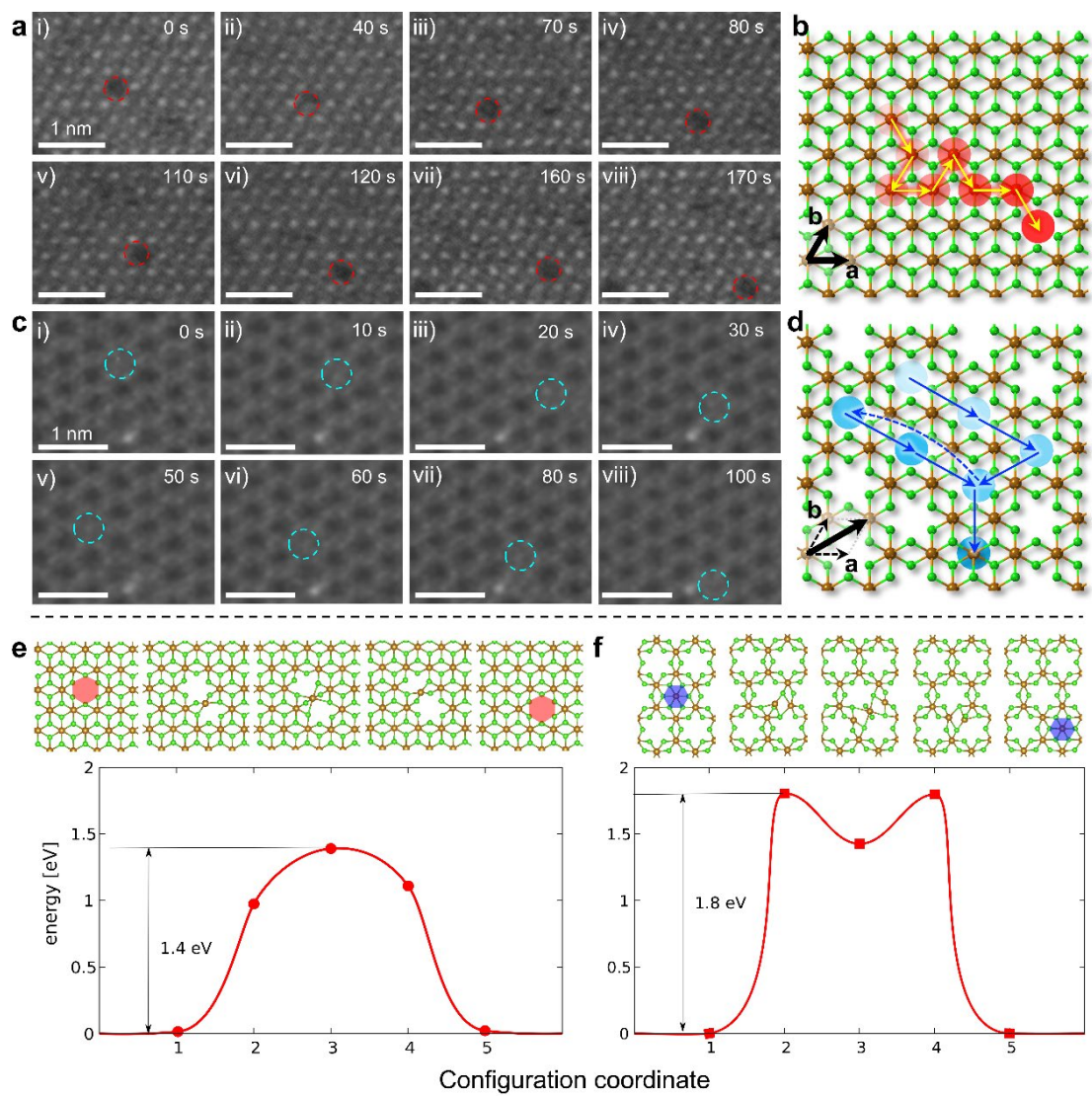

**Figure 2. Migration behavior of Fe vacancies and Fe interstitials.** (a) A sequence of ADF images showing the migration of a single Fe vacancy, marked by a red dashed circle, in $FeCl_2$. (b) Atomic models illustrating the migration steps of the Fe vacancy. (c) Sequential ADF images depicting the migration of a single Fe interstitial, highlighted by cyan dashed circles, in $FeCl_3$. (d) A schematic model showing the migration trajectory of a single Fe interstitial. (e) The energy profile and intermediate atomic configurations of a Fe vacancy migrating in the $FeCl_2$ lattice indicate a migration barrier of 1.4 eV. (f) The energy profile and intermediate atomic configurations of a Fe interstitial migrating in $FeCl_3$ lattice indicate a migration barrier of 1.8 eV.

**Migration behavior of multiple Fe vacancies in FeCl$_2$**

Isolated Fe vacancies can migrate randomly within the FeCl$_2$ domain when energy is supplied by impinging accelerated electrons. However, our experiment often observed that Fe vacancies tend to form divacancies, as shown in **Figure 3a** (see also Movie S4). The divacancy, marked by yellow dashed circles, remained intact even as it changed locations. Interestingly, the divacancy sometimes split into single vacancies but rejoined later on, as illustrated in the 170-second and the subsequent frames. This observation is supported by our DFT calculations, which indicate that divacancies are energetically more favorable as compared to two isolated Fe vacancies. The energy difference between these configurations ranges from 1.8 to 4 eV, depending on the location of the Fe vacancies. Additionally, the calculations reveal that it is energetically unfavorable to release the two Cl atoms situated between the two missing Fe atoms in the divacancy. Releasing a Cl$_2$ molecule requires approximately 1 eV, ensuring that the FeCl$_2$ lattice remains intact even as Fe vacancies migrate within the lattice (Figure 3b). Overall, the energy of the FeCl$_2$ phase (per structural unit) plus half the energy of isolated Cl$_2$ molecule is higher than that of FeCl$_2$ by 0.5 eV. Moreover, the migration path of isolated Fe vacancies in FeCl$_2$ is limited to a few nanometers, and prolonged scanning does not significantly alter the overall structure or vacancy density.

Interestingly, when the number of connected Fe vacancies increases, forming small domains of FeCl$_3$ within the FeCl$_2$ lattice, their migration behavior differs from that of single vacancies. Figure 3c presents sequential ADF images showing the movement of Fe vacancy groups, specifically several small domains of FeCl$_3$. These groups of Fe vacancies can migrate collectively over longer distances as compared to isolated Fe vacancies in FeCl$_2$ lattice. However, the migration of Fe vacancy groups does not expand the FeCl$_3$ domain size sufficiently to induce a phase transformation from FeCl$_2$ to FeCl$_3$ under electron beam scanning (see Movie S5). On the contrary, we observed that electron beam-induced phase transformations are more likely to occur from FeCl$_3$ to FeCl$_2$, a topic that will be discussed in the next section.

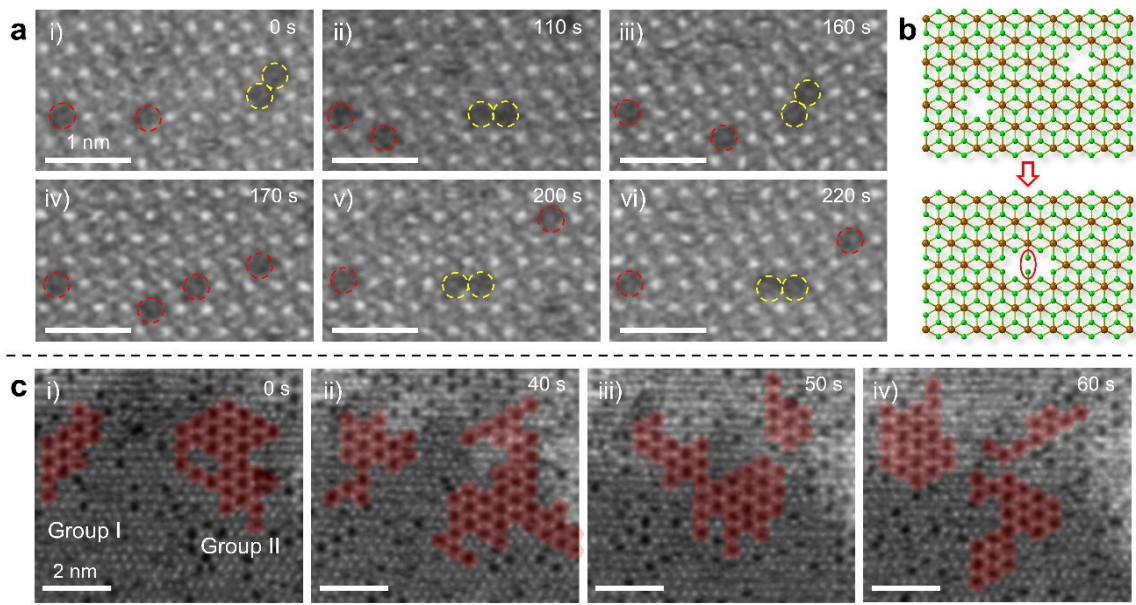

**Figure 3. Dynamics of multiple Fe vacancies in FeCl$_2$.** (a) Selected frames from a sequence of ADF images showing two joint Fe vacancies forming a divacancy, highlighted by yellow dashed circles. (b) Atomic models illustrating the FeCl$_2$ structure with two separated Fe vacancies and a Fe divacancy. (c) Groups of Fe vacancies forming FeCl$_3$ nanodomains within the FeCl$_2$ lattice, highlighted in red and labeled as Group 1 and Group 2 in frame (i). The shape of these FeCl$_3$ nanodomains changes dramatically in each subsequent frame, indicating that the Fe vacancy groups move collectively.

## Phase transformation from FeCl$_3$ to FeCl$_2$

In the previous section, we established that FeCl$_2$ does not transform into FeCl$_3$ through the migration of Fe vacancies, even though this is energetically favorable. Structurally, the FeCl$_3$ lattice contains 33% more "Fe vacancies" as compared to FeCl$_2$, making FeCl$_2$ a more closely packed structure. Increasing the number of Fe vacancies in $FeCl_2$ would require Fe atoms to be displaced or migrate outwards to form the periodic $FeCl_3$ lattice, both scenarios are highly unlikely.

Conversely, the transformation from $FeCl_3$ to $FeCl_2$ is more probable and was frequently observed during our STEM experiments. The *in-situ* phase transformation from $FeCl_3$ to $FeCl_2$ was recorded and presented in **Figure 4a** and Movie S6. $FeCl_2$ domains, highlighted in blue, grow from the edges of the imaging area where the electron dose is higher. Our DFT calculations indicate that Cl atoms are more likely to be sputtered by the e-beam than Fe atoms (Figure 4b). This sputtering of Cl atoms at the imaging corners facilitates the migration of Fe atoms toward the imaging center, filling in the Fe vacancies of $FeCl_3$ and thus forming a denser $FeCl_2$ lattice without changing the total number of Fe atoms. We note that, as discussed above, transformation of $FeCl_3$ to $FeCl_2$ with release of chlorin molecules is energetically favorable, but not the reverse transformation. However, the irradiation-induced transformations cannot be explained by simple energy balance. The system is dissipative and not in thermal equilibrium. Most important, Cl atoms are sputtered more easily from $FeCl_3$ than from $FeCl_2$. The displacement cross-section for a Cl atom at 60 kV is three times higher for $FeCl_3$ (210 barn) than for $FeCl_2$ (70 barn) according to our calculations. Based on the cross section values one can say that electron irradiation drives the system into a more radiation-resistant state, similar to the behavior of carbon systems under electron beam: transformations of graphitic structures into diamond have been observed in TEM,[29, 30] although the latter has higher energy than the former. This was explained by a lower cross section for the displacement of carbon atoms in diamond, that is a higher stability of $sp^3$-hybridized carbon under irradiation as compared to the $sp^2$ systems.[31]

The corresponding phase transformation process is illustrated in the schematic model shown in Figure 4c. Under continuous electron beam scanning, Fe atoms migrate collectively along directions marked by blue arrows, generating small isolated $FeCl_2$ domains that eventually merge into a larger $FeCl_2$ domain. The movement of Fe adatoms (Figure 1k) does not contribute to the phase transition from $FeCl_3$ to $FeCl_2$. As shown in Figure 1a, Fe adatoms are observed exclusively in $FeCl_3$ and not in $FeCl_2$. This is because Fe adatoms migrate outside of $FeCl_3$ lattice and do not participate in the internal Fe migration for phase transformation. Additionally, under electron beam scanning, Fe adatoms exhibit greater stability compared to Fe vacancies in $FeCl_2$ and Fe interstitials in $FeCl_3$ (as shown in Figure 2). Based on this evidence, it can be concluded that Fe adatoms on the lattice do not contribute to the phase transition between $FeCl_2$ and $FeCl_3$. Notably, this electron beam-induced phase transformation occurs only in the irradiated region (Figure S5), allowing for the patterning of $FeCl_2$ islands within the $FeCl_3$ domain at specific locations.

Building on these findings, we further examined phase transformation behaviors in a bilayer $FeCl_3$ intercalated region to explore thickness-dependent phase transition process. Our study system mainly consists of over 99% single-layer $FeCl_3$ intercalated in BLG. However, a bilayer $FeCl_3$ intercalated region was identified, and the corresponding results are presented in Figure S6. In this region, the second $FeCl_3$ layer stacks with a twist angle of 6.3°, forming a moiré pattern. After electron beam scanning for 100 seconds, a phase transformation was observed, but it occurred only in the larger continuous single-layer region. It is understood that the $FeCl_3$-to-$FeCl_2$ transition requires the sputtering of Cl atoms and the spontaneous migration of Fe atoms. In the bilayer region, one of the layers was effectively protected, resulting in the phase transformation occurring exclusively in the single layer.

We further investigated theoretically how the electronic properties of $FeCl_x$ are influenced by phase transformations. The $FeCl_2$ monolayer exhibits a half-metallic ground state, as evidenced by its electronic structure (Figure S7), in agreement with the results of previous calculations.[32, 33] The spin-polarized density of states clearly shows that the minority spin states from Fe atoms cross the Fermi level, while the majority spin states exhibit a significant energy gap. In contrast, bulk $FeCl_2$ displays an indirect band gap and an antiferromagnetic (AFM) magnetic state. For $FeCl_3$, the half-metallic nature is preserved in both the monolayer and bulk systems. The presence of Fe vacancy modifies the electronic structures of $FeCl_2$ monolayer by introducing new spin-polarized states around the Fermi level as shown in Figure S8.

We also studied the electronic structure of $FeCl_x$ monolayers intercalated between bilayer graphene (BLG). Based on our charge density difference plots (Figure S9), which depict electron excess and depletion with blue and red areas, we observed electron transfer from graphene to the $FeCl_2$ and $FeCl_3$ layers, leading to n-type doping of the intercalant. Our findings align with previously reported data for $FeCl_3$.[34] The amount of charge transferred from graphene to $FeCl_x$ was further examined using Bader analysis (Table S1). The results showed a higher electron density transferred from graphene to $FeCl_2$ as compared to $FeCl_3$. The presence of Fe vacancy in $FeCl_2$ further enhances the charge transfer in comparison to pristine material, as shown in Figure S10.

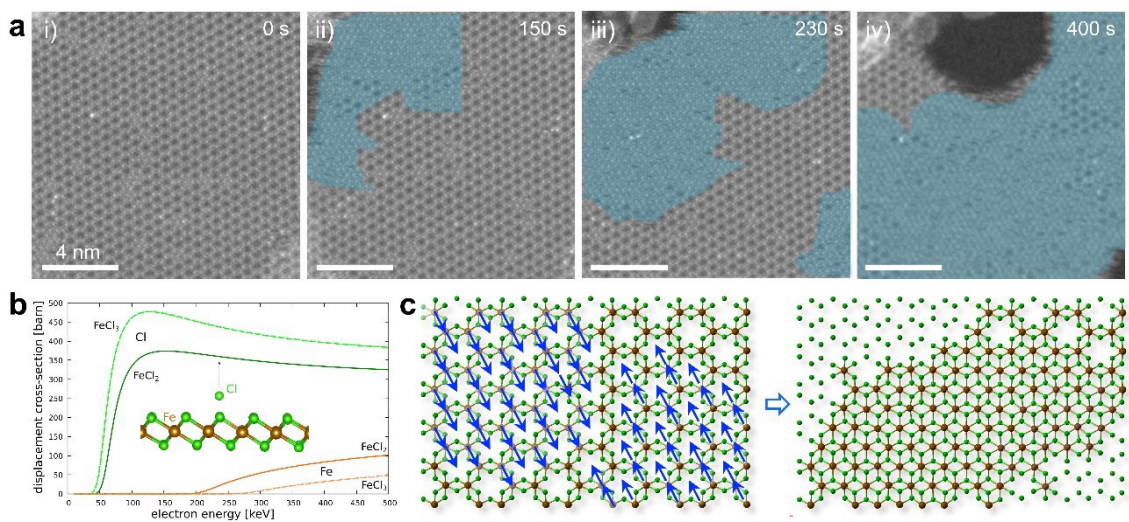

**Figure 4. Migration of Fe interstitials in FeCl₃ and phase transformation.** (a) Sequential ADF images reveal the structural transformation from $FeCl_3$ to $FeCl_2$. The $FeCl_2$ domains, indicated by blue shading, grow from the borders of the imaging area where the electron dose accumulates. (b) Displacement cross-section threshold for Cl and Fe atoms, calculated using the McKinley-Feshbach formalism, showing the susceptibility of Cl atoms to sputtering under electron beam irradiation. (c) Schematic atomic models illustrating the phase transformation process from $FeCl_3$ to $FeCl_2$. The collective migration of Fe atoms, indicated by blue arrows, fills the central Fe vacancies in $FeCl_3$, leading to the formation of $FeCl_2$ domains.

## Unexpected iron chloride and iron oxychloride nanostructures

When examining the intercalated structures of $FeCl_2$ and $FeCl_3$ in BLG, we discovered a unique nanostructure with a stoichiometry of $Fe_5Cl_{18}$, distinct from the standard iron chloride phases, which, to the best of our knowledge, has not been observed before (**Figure 5a**). The crystallinity of this structure presented in Figure 5a is also evident from the filtered inverse fast Fourier transform (IFFT) image in Figure 5b, and the DFT-simulated atomic model is shown in Figure 5c. We note that the proposed atomic model (calculated without graphene and at zero temperature) is scaled up by approximately 18% to align with the experimental dimensions. This lattice mismatch may be attributed to the high-pressure environment within the interlayer space of BLG, charge transfer from/to graphene and finite temperature in the experiment. The pressure created by the graphene sheets can be anisotropic, that is acting in different directions depending on the specific location. In the perfectly flat geometry, one can expect that pressure applied perpendicular to the material may "squash" it and make laterally larger, by analogy of a piece of rubber in a mechanical vice. Next to the edge of the encapsulated material, where graphene sheets meet, there may be a component of the force acting parallel to the sheets. We assume that the first scenario is relevant. We would like to emphasize that pressure could be only one of the reasons, charge transfer and finite temperatures can be even more important. Remarkably, this $Fe_5Cl_{18}$ nanostructure can be crystallized from amorphous iron chloride under electron beam irradiation (Figure 5d-f and Movie S7). Under continued e-beam exposure, it further transforms into $FeCl_2$ as chlorine atoms are progressively displaced, as shown in Figure 5g-i (See Movie S8).

A region with brighter contrast in the lower-left side of Figure 5f indicates the presence of iron oxychloride (FeClO), an oxidized phase induced by electron beam exposure. This transformation typically initiates at the edges of the $FeCl_3$ domains (Figure 5j-m and Movie S9), where the edges first become amorphous, react with oxygen, and subsequently recrystallize as FeClO, expanding until the local oxygen source is depleted. This oxygen likely originates from the adsorbed hydrocarbons or impurities from the intercalant source, as $FeCl_3$ readily hydrates upon exposure to air or moisture. A summary of all identified intercalated iron chloride nanostructures and their corresponding spectra is provided in Figure S11. The quantitative EELS data, presented in Figure S12, confirms the presence of $FeCl_2$, $FeCl_3$ and $Fe_5Cl_{18}$ phases based on the calculated Fe-to-Cl ratios. These findings further underscore the unique interlayer spacing of BLG as an environment that stabilizes novel 2D materials, enables detailed investigation of defect structures, and even functions as a nanoscale chemical reaction cell for in-situ observations.

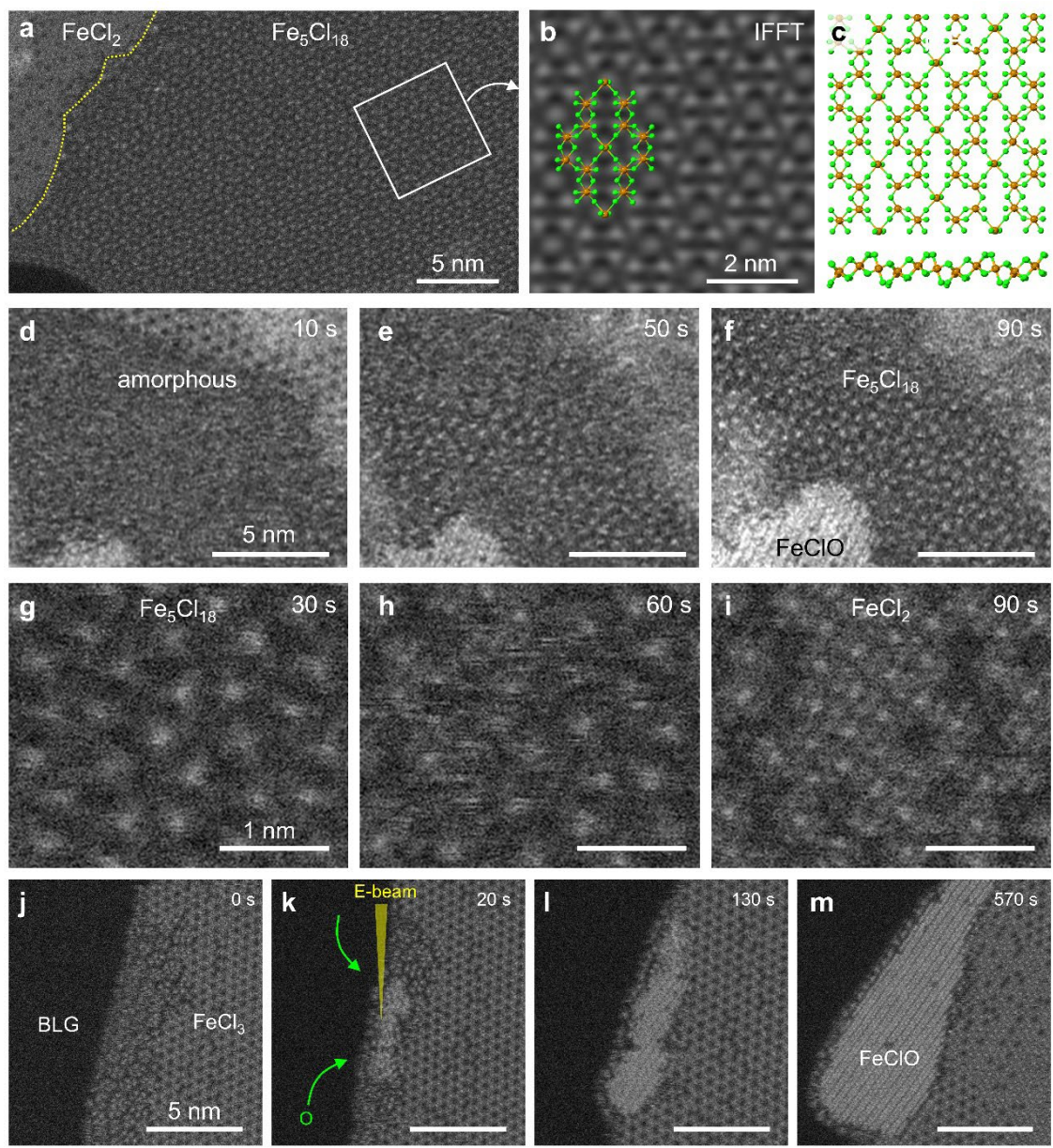

**Figure 5. Structural dynamics of Fe$_5$Cl$_{18}$ and FeClO in BLG.** (a) ADF image of Fe$_5$Cl$_{18}$ intercalated in BLG. (b,c) IFFT image and atomic model of the Fe$_5$Cl$_{18}$. (d-f) Sequential ADF images depicting the crystallization process of Fe$_5$Cl$_{18}$ initiated by e-beam irradiation. (g-i) ADF images showing the transformation of Fe$_5$Cl$_{18}$ into FeCl$_2$ over time. (j-m) Sequential ADF images illustrating the e-beam-induced growth of FeClO from the edge of FeCl$_3$.

## Conclusions

By combining STEM experiments and first-principles calculations, we studied the behavior of atomic-scale defects in iron chlorides intercalated into BLG and elucidated their role in the transformations between the FeCl$_2$ and FeCl$_3$ phases we observed. We provided insights into the atomic mechanisms driving these transformations in terms of defect diffusivity, as well as the energetics of the phases and their stability under electron beam. A crystalline phase with an unusual stoichiometry of Fe$_5$Cl$_{18}$ which has not been reported before was also observed. We further investigated the impact of atomic-scale defects on the electronic structure and properties of iron chlorides. The patterning of local FeCl$_2$ domains within the FeCl$_3$ lattice demonstrates the potential to modify magnetic[35] properties through targeted electron beam manipulation. Our findings enhance our understanding of the behavior of intercalated structures with defects and their effects on the performance and stability of such materials in practical applications.

## Method

### FeCl$_3$ intercalation

For the FeCl$_3$ intercalation into BLG, we used a method similar to the one previously reported for metal chloride intercalation in BLG.[4-6] BLG grown by chemical vapor deposition on Cu foils[36] and Cu-Ni/sapphire substrates[37, 38] were used for the FeCl$_3$ intercalation. BLG was first transferred onto a TEM grid. The transfer method was as following: A layer of polycarbonate (2.5 wt% dissolved in chloroform) was coated to the BLG grown on Cu foil or on NiCu alloy, which was then placed in an aqueous hydrochloric acid solution to etch out the Cu foil and then thoroughly cleaned with pure water before the BLG sheet was transferred to a TEM grid, then the polycarbonate film on the BLG surface was washed with the standard cleaning steps.[22] Next, FeCl$_3$ powder and BLG-TEM grid were encapsulated in a glass tube, which was then pumped to a vacuum below 10$^{-3}$ Pa and transferred to a box furnace heated at 250°C for 12 h. Then the glass tube was opened in a glove box, and the TEM grid was transferred to a JEOL vacuum transfer holder to avoid possible degradation under ambient conditions.

**STEM and EELS**

STEM images were acquired with JEOL TripleC#3, an ARM200F-based ultra-high vacuum microscope equipped with a JEOL delta corrector and a cold field emission gun operating at 60 kV. The ADF probe current was ~10 pA, and the convergence half-angle and internal acquisition half-angle were 37 mrad and 76 mrad, respectively. The resolution of a typical ADF image was 1024 ×1024 pixels. The EELS core loss spectrum was acquired by line scan with an exposure time of 0.1 s/pixel and recorded with a Gatan Rio CMOS camera optimized for low voltage operation. All STEM images and EELS spectra were acquired at room temperature.

**Density functional theory calculations**

Spin-polarized simulations were performed using the Vienna ab-initio Simulation Package (VASP).[39, 40] The structures were relaxed using the Perdew–Burke–Ernzerhof (PBE) exchange-correlation functional,[41] with a force tolerance of 0.01 eV/Å and electronic convergence criteria of $10^{-5}$ eV. Brillouin zone integration with the tetrahedron method and Blöchl corrections were applied with 5×5×1 k-points. vdWs interactions were considered using the DFT-D3 method with Becke-Jonson damping.[42] The nudged elastic band calculations were performed using five images to determine the transition barrier between the possible $FeCl_x$ configurations. Furthermore, the structures of the chosen phases were analyzed using an advanced TS dispersion correction method, which includes the iterative Hirshfeld partitioning scheme.[43]

The simulation models of $FeCl_x$ sandwiched between graphene bilayer were constructed using 5x5 graphene combined with 2x2 $FeCl_2$ slab and 7x7 graphene with 3x3 $FeCl_3$ slab. The Brillouin zone of the primitive cells and supercells were sampled using 12x12x1 and 4x4x1 k-points, respectively. The VESTA package was used for visualizing the charge density differences.[44]

## Data availability

The data that support the plots within this paper and other finding of this study are either provided in the Article and its Supplementary Information or available from the corresponding author upon request.

## Acknowledgment

This work was financially supported by the JSPS Grant-in-Aid for Scientific Research on Innovative Areas "Science of 2.5 Dimensional Materials: Paradigm Shift of Materials Science Toward Future Social Innovation" Y.C.L acknowledge to JSPS-KAKENHI (22H05478). H.A. acknowledge to JSPS-KAKENHI (18H03864, 19K22113, 21H05233). K.S. acknowledge to JSPS-KAKENHI (16H06333, 21H05235, 22F22358), the JST-CREST program (JPMJCR20B1, JMJCR20B5, JPMJCR1993), ERC "MORE-TEM (951215), and the JSPS A3 Foresight Program.

A.V.K. acknowledges funding from the German Research Foundation (DFG), projects KR 4866/9-1, and the collaborative research center "Chemistry of Synthetic 2D Materials" CRC-1415-417590517. Generous CPU time grants from the Technical University of Dresden computing cluster (TAURUS) and Gauss Centre for Supercomputing e.V. (www.gauss-centre.eu), Supercomputer HAWK at Höchstleistungsrechenzentrum Stuttgart (www.hlrs.de), are greatly appreciated.

P.W.C. acknowledge to the Ministry of Science and Technology (MOST) Taiwan Grants MOST 109-2124-M-007-002-MY3, MOST 109-2112-M-007-027-MY3, MOST 106-2628-M-007-003-MY3, and MOST 109-2124-M-006-001 as well as Academia Sinica (AS) Grant AS-TP-106-A07.

## Author contributions

Y.C.L. and K.S. supervised the project. Y.C.L., Q.N.L and H.M.S. designed and performed the STEM experiments. S.K., M.G.-A. and A.V.K. performed the DFT and

MD calculations. C.C.C., P.W.C., P.S.F., and H.A. synthesized the bilayer graphene specimen. All authors contributed to the discussion of the project. Q.N.L. analyzed the data. Q.N.L., Y.C.L., A.V.K., and K.S. wrote the manuscript with input from all co-authors.

## Competing interests

The authors declare no competing interests.

## Additional information

**Supplementary information** The online version contains supplementary material available at https://doi.org/xxxxxxx.

**Correspondence** and requests for materials should be addressed to Yung-Chang Lin.